\documentclass[aps,twocolumn,prl,superscriptaddress,showpacs,preprintnumbers,amsmath,amssymb]{revtex4}

\usepackage{graphicx}
\usepackage{color}

\begin{document}

\title{Symmetry Enforced Dirac Points in Antiferromagnetic Semiconductors}
\author{V.~V.~Kabanov}
\affiliation{Jozef Stefan Institute, Jamova 39, 1000 Ljubljana, Slovenia}

\date{\today}

\begin{abstract}
It is shown that the symmetry enforced Dirac points exist at some time reversal symmetric momenta in antiferroemgnetic compound GdB$_4$. These Dirac points may be controlled by the external magnetic field or by the deformation of the crystal. Application of the external magnetic field leads to splitting of these points into Weyl points or to opening of a gap depending on the field direction. The application of the symmetry breaking deformation also opens a gap in the spectrum. Suppression of the antiferromagnetic order leads to the formation of the nodal line instead of the Dirac points. This indicates that the symmetry enforced Dirac semimetals may be effectively used in different spintronic devices.  
\end{abstract}


\maketitle
\section{Introduction}
Recent developments in the field of antiferromagnetic spintronics\cite{MacDonald,Baltz,Jungwirth} have stimulated the search for materials susceptible to  the antirerromagnetic (AFM) order parameter. There are few reports where the AFM order was controlled by an electric current in CuMnAs \cite{Wadley,Smejkal} and in Mn$_2$A \cite{Bodnar}. Among the materials which are very susceptible to the AFM order are Dirac semimetals. It was argued that the room temperature AFM metal MnPd$_2$ allows the electric control of the Dirac nodal line \cite{Shao}. Dirac fermions were predicted in AFM semimetals  CuMnAs and CuMnP \cite{Tang}. In these materials it was suggested that the Dirac fermions may be electrically controlled by the spin-orbit torque \cite{Smejkal}. 

It is known that there are two types of Dirac semimetals \cite{Armitage}. The first type occurs due to band inversion. In that case the Dirac node can occur when two inverted bands undergo an accidental band crossing. This crossing is unstable at any general point of the Brillouin zone (BZ). For a general point of the BZ the small group is trivial and these two bands will be hybridized producing a band gap \cite{Armitage}. The situation may be different when the crossing occurs along some high symmetry line in the BZ. In that case the two crossing bands may belong to two different irreducible representations of the small group and this prevents the hybridization \cite{Armitage}. This type of Dirac points is very sensitive to parameters of the Hamiltonian. It means that by continuous tuning of the parameters of the Hamiltonian we may uninvert the bands and two symmetric Dirac points annihilate \cite{Armitage}. Note that all proposed Dirac points in AFM metals \cite{Smejkal,Shao,Tang} belong to this type. It is clear from the Hamiltonian introduced in Ref.\cite{Smejkal} (Eq.(1) in Ref.\cite{Smejkal}), that if we continuously change the exchange interaction and make it larger than the spin-orbit coupling the two Dirac points disappear without any change of the symmetry. 

The second type of Dirac semimetals is the symmetry enforced Dirac semimetal \cite{Armitage,Young}. This type of a Dirac point does not depend on the parameters of the Hamiltonian and is determined only by the symmetry properties of the system. The symmetry criterion for the existence of the symmetry enforced Dirac semimetal was formulated by S.M. Young et.al. \cite{Young}. The group should allow four dimensional spinor representations of the small group at some point \textbf{k} of the BZ. The band velocities must be nonzero  at this point \textbf{k}. It means that the square of the four dimensional spinor representation of the small group must contain the vector representation \cite{BirPikus}. And finally, branches of the valence and the conduction bands should not be degenerate away from \textbf{k}. There are substantial amount of the space groups that satisfies to these criteria \cite{Armitage}. Note that the only way to destroy the symmetry enforced Dirac point is to reduce the symmetry. Any continuous change of the parameters of the Hamiltonian will not affect this Dirac point. 

The situation in AFM semimetals is different, because the symmetry of an antiferromagnet is substantially different from the symmetry of a paramagnetic semimetal. In AFM semimetals the time reversal symmetry is broken. It means that in general the Kramers degeneracy is broken. Nevertheless, in AFM materials the time reversal operation $\theta$ very often comes together with some spatial operation and this leads to the restoration of the Kramers degeneracy. This situation occurs in CuMnAs \cite{Smejkal,Tang}, where the product of the time reversal operation 
$\theta$ with the inversion symmetry $I$ is the true symmetry operation of the AFM crystal. 

In order to describe the symmetry enforced Dirac points in AFM materials we have to know  irreducible corepresentations of the nonunitary group \cite{Kovalev,Bradley}. The criterion for the symmetry enforced Dirac point is almost the same as in the paramagnetic semimetal. The nonunitary group should allow four dimensional spinor corepresentations of the small group at some point \textbf{k} of the BZ, the band velocities must be nonzero at this point, and this degeneracy should be lifted away from the \textbf{k}-point. 

In addition there is one very interesting and very specific property of AFM conductors. Usually application of the magnetic field in ordinary metals and semiconductors lifts the Kramers degeneracy. In AFM conductors there are special points in the BZ where the Kramers degeneracy is preserved when the magnetic field is perpendicular to the sublattice magnetization of the AFM conductor \cite{Ramaz1,Ramaz2}. This is usually observed experimentally as the absence of g-factors in quantum oscillation measurements \cite{Kabanov}. This is again related to the fact that the spin rotation together with the time reversal operation and the AFM translation is the true symmetry operation in the AFM crystal \cite{Ramaz1,Ramaz2}.

In this paper using AFM crystal GdB$_4$ as an example, I show that the symmetry enforced Dirac points exist at some high symmetry points of the BZ. These Dirac points are very sensitive to the symmetry and they disappear in the paramagnetic phase forming the Dirac nodal line. Application of the external magnetic field splits a Dirac point into two Weil points with opposite Chern numbers when the field is perpendicular to magnetization in all sublattices. At certain directions of the external magnetic field the Kramers degeneracy is preserved at some high symmetry points of the BZ even in the case of strong sin-orbit interaction and in the case of noncollinear AFM order. Here I assume that the spin-orbit interaction is strong and spins are transformed by lattice rotations, i.e. the rotational symmetry in the spin space is broken.

\begin{figure}[tb]
\includegraphics[angle=0,width=1.0\linewidth]{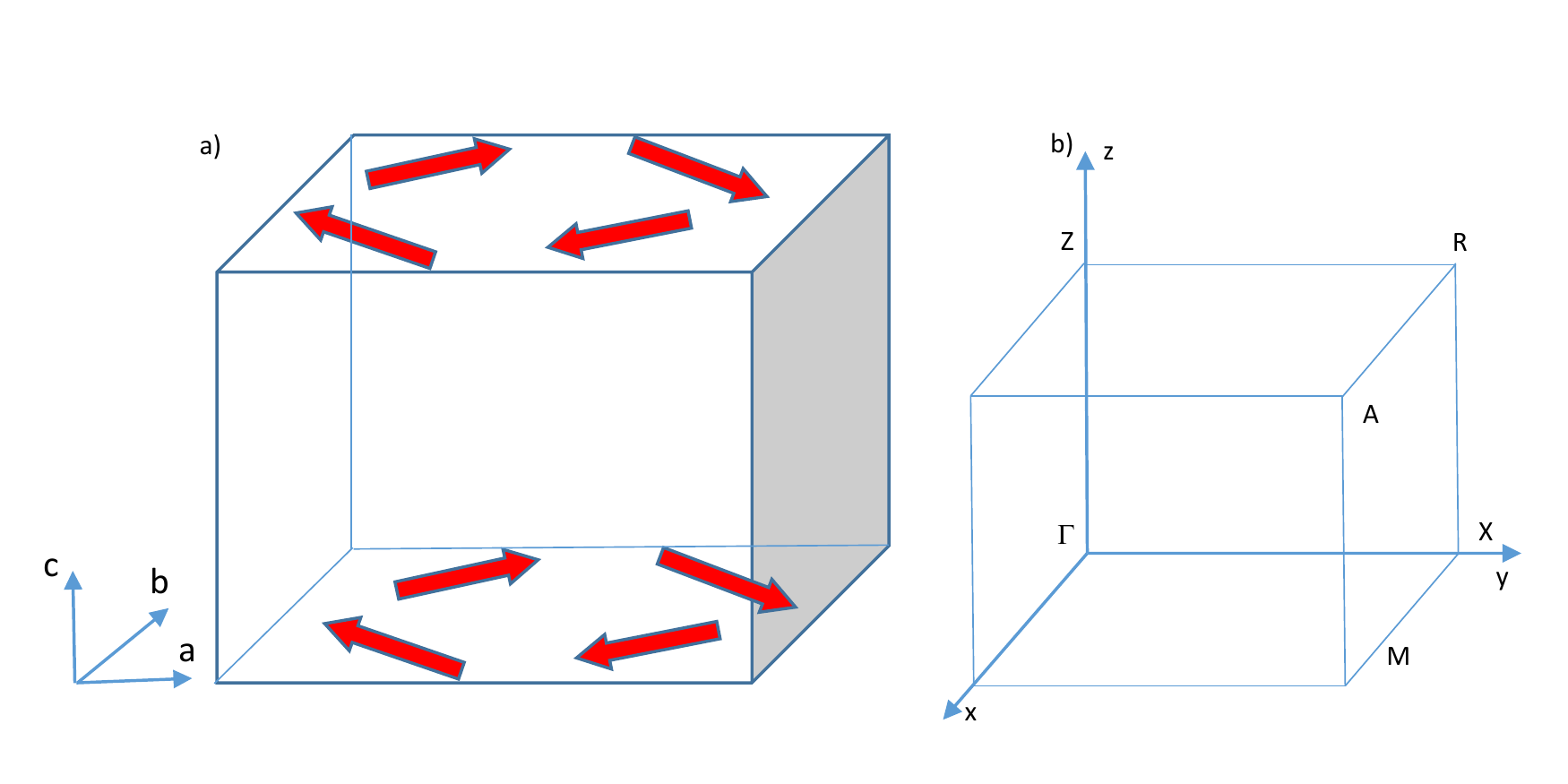}
\caption{a) The magnetic structure of the AFM state of the tetragonal GdB$_4$. Arrows indicate the direction of the magnetic moments of Gd ions. b) High symmetry TRIM in the BZ of GdB$_4$. $R$ and $X$ points have the two arms costar, $\Gamma$, $Z$, $M$ and $A$ points have the one arm costar.}
\label{fig:AC}
\end{figure}

\section{Theory of a symmetry enforced Dirac point in Antiferromagnets}
Fig. 1a shows the crystal structure of tetragonal GdB$_4$ in the metallic \cite{Pluzhnikov} AFM state. The Gd atoms are arranged in four sublattices AFM structure with magnetic moments $\mathbf{M}=(\pm M, \pm M,0)$ with $M=5.05\mu_B$, $\mu_B$ is the Bohr magneton. Above the Neel temperature T$_n$=42 K in paramagnetic phase it has a nonsymmorphic space group $P4/mbm$ (\#127). Below the Neel temperature T$_N$ it has magnetic space group $P4/m'b'm'$ (\#127.395) in Belov-Neronova-Smirnova settings \cite{Bradley}. The nonunitary group for this material may be written as
\begin{equation}
\mathbf{G}=\mathbf{D_4^2}+\theta I\mathbf{D_4^2},
\end{equation}
here $\theta$ is the time reversal antiunitary operation, $I$ is the space inversion and 
$\mathbf{D_4^2}$ (\#90) is the unitary subgroup of the magnetic space group. It is important to underline that for GdB$_4$ neither the inversion $I$ or the time inversion $\theta$ are the symmetry operation, but the product $\theta I$ is the true symmetry operation of the AFM crystal. It means that the Kramers degeneracy is preserved. 

The standard way of construction of corepresentations is as follows. First the time reversal invariant momenta (TRIM) of the BZ should be identified. For this magnetic group there are following TRIM $\Gamma$($0,0,0$), $X$($0,\pi/\tau,0$), $R$($0,\pi/\tau,\pi/\tau_z$), $M$($\pi/\tau,\pi/\tau,0$), $Z$($0,0,\pi/\tau_z$) and $A$($\pi/\tau,\pi/\tau,\pi/\tau_z$),(Fig.1b) where $\tau$ is the translation in x and y direction and $\tau_z$ is translation along z axis. $X$ and $R$ points have the two arms costar all other momenta of TRIM have the single arm costar. $\Gamma$ point is not relevant because it does not have four dimensional corepresentations. For all other points it is necessary to conduct further analysis. The next step is to identify the the small group and the unitary subgroup for every TRIM. After the small group and unitary subgroup are identified the irreducible representations of the unitary subgroup should be constructed \cite{Kovalev}. In order to construct irreducible corepresentations of the small group from irreducible representations of the unitary subgroup the type of the corepresentation should be identified.   For that purposes the sum of characters of squares of the antiunitary elements should be calculated
\begin{equation}
\sigma=\frac{1}{N} \sum_a \chi(a^2), 
\end{equation}  
where $N$ is the order of the unitary subgroup \cite{Bradley}. 
If $\sigma=1$, the corepresentation belongs to type "a" and the irreducible representation of the unitary subgroup generates one irreducible corepresentation of the small group of the same dimensionality. There is no additional degeneracy in that case. If $\sigma=-1$ the corepresentation belongs to the type "b" and the irreducible representation of the unitary subgroup generates one irreducible corepresentation of doubled dimensionality. Therefore, there is additional degeneracy in that case. And finally, if $\sigma=0$ the corepresentation belongs to type "c". In that case the two irreducible representations of the unitary subgroup with $\sigma=0$ are merged to form one corepresentation with doubled dimensionality. This also leads to additional degeneracy. The matrices of irreducible corepresentations may be easily constructed using the formulas from \cite{Kovalev,Bradley}. 

At this point it is easy to see that the Kramers degeneracy is preserved in GdB$_4$. In a general point of the BZ the smal group contains only two elements: $e$ and $\theta I$. There is only the trivial representation of the unitary subgroup of the small group. The square of the antiunitary element $(\theta I)^2=-e$. Here the relation $\theta^2=-e$ for spinor representations was used \cite{Bradley}. Therefore, $\sigma=-1$ and the one dimensional representation of the unitary subgroup generates the two dimensional corepresentation of the small group. Therefore, the Kramers degeneracy is preserved. 

The small groups for $X$ and $R$ points coincide and consist of $e, c_{2x}|\alpha), c_{2y}|\alpha), c_{2z}, \theta I, \theta\sigma_x|\alpha), \theta\sigma_y|\alpha), \theta\sigma_z$, where $\alpha=(\tau/2,\tau/2,0)$ is nontrivial translation, and $N=4$. There are 4 one dimensional irreducible representations of the unitary subgroup of the small group   \cite{Kovalev}. The sum of characters of squares of the antinunitary elements
(Eq.(2)) $\sigma=0$
for both $X$ and $R$ momenta. Therefore, these corepresentations belong to type "c" \cite{Kovalev,Bradley}. As a result, two two-dimensional corepresentations of the small group are formed. Therefore, we conclude that $X$ and $R$ points cannot host the Dirac point. 

The small group for TRIM $Z$, $M$, and $A$ coincides with the symmetry group of the AFM state (Eq.(1)) with $N=8$. The main difference between $Z$ point and $M$ and $A$ points is that the phase factors due to nontrivial translations are different. In all cases the unitary subgroup of the small group has two two-dimensional irreducible representations \cite{Kovalev}. Since for the momentum $Z$($0,0,\pi/\tau_z$) the exponent $\exp{(-2i\mathbf{k}\alpha)}=1$, the nontrivial translations do not have any effect on the criterion Eq.(2). The situation is exactly the same as in $\Gamma$ point. Therefore, $\sigma=1$, 
and all irreducible corepresentations belong to type "a" \cite{Kovalev,Bradley}. Therefore, each irreducible representation of the unitary subgroup generate one irreducible corepresentation of the same dimensionality. Therefore, in $Z$ point of the BZ there are two two-dimensional corepresentations. This point cannot host a Dirac point as well. 

The situation is very different at TRIM $M$ and $A$. The unitary subgroup of the small group has two two-dimensional representations, which formally are equal for both TRIM \cite{Kovalev}.   
But because $\exp{(-2i\mathbf{k}\alpha)}=-1$ for both TRIM, the sum of squares of the antinunitary elements $\sigma=0$.
Therefore, these two representations belong to type "c" \cite{Kovalev,Bradley} and form the four-dimensional corepresentation. Therefore, TRIM $M$ and $A$ can host a Dirac point. Indeed, the square of the corepresentation at $M$ and $A$ points contains the vector representation and the four fold degeneracy is lifted when $\mathbf{k}$ moves away from $M$ and $A$ points of the BZ.

Note that the situation is different in the paramagnetic phase of GdB$_4$. Again, the corepresentations in the points $M$ and $A$ are four dimensional. Therefore, these points can host a Dirac point. Nevertheless, this degeneracy is not lifted along $M-A$ (Fig.1b) line of the BZ. Therefore, these points do not host the true Dirac points. This fact demonstrates that the details of the spectrum may be controlled by the destruction of the AFM order.

In order to construct the Hamiltonian describing the spectrum near $M$ and $A$ points it is necessary to construct the matrices of irreducible spinor corepresentations for the group generators. Since all irreducible corepresentations belong to type "c" this construction is reduced to merging of two two-dimensional representations of the unitary subgroup in one four dimensional corepresentation \cite{Kovalev,Bradley}.
Using data from Ref. \cite{Kovalev}, one can construct irreducible corepresentations for group generators  $D(c_{2x}|\alpha))=-\tau_0\otimes\sigma_z$, $D(c_{4z})=-\tau_z\otimes (i\sigma_0+\sigma_y)/\sqrt{2}$, and $D(\theta I)=-\tau_y\otimes i\sigma_0$, where $\tau_i$ and $\sigma_j$ are two sets of Pauli matrices. The irreducible corepresentations for other elements of the group may be obtained from the products of these generators. 

Consider the Hamiltonian matrix constructed in the same basis as irreducible corepresentation. This Hamiltonian transforms under group elements $g$ as:
\begin{equation}
D^{-1}(g)H(\mathbf{k},\mathbf{H})D(g)=H(g^{-1}\mathbf{k},g^{-1}\mathbf{H})
\end{equation}  
Here $\mathbf{k}$ is the deviation of momentum from $M$ or $A$ point, and $\mathbf{H}$ is external magnetic field. This relation expresses the invariance of the Hamiltonian towards the transformation $g$ \cite{BirPikus} (see also Ref.\cite{Manes}) This equation may be used to construct Hamiltonian matrix. In practice it is useful to consider 16 basis $4\times 4$ matrices $\tau_i\otimes\sigma_j$, and construct invariant forms using this set of matrices.  

Using matrices of corepresentations for the group generators it is easy to show that the following   expressions are invariant under the group transformation $\tau_0\otimes(k_x\sigma_z+k_y\sigma_x)$ and $k_z\tau_z\otimes\sigma_y$. Therefore, the Hamiltonian has the following form:
\begin{equation}
H_k=v_1\tau_0\otimes(k_x\sigma_z+k_y\sigma_x)+v_2k_z\tau_z\otimes\sigma_y
\end{equation}
This Hamiltonian has two independent velocities $v_1$ and $v_2$ and has the standard Dirac spectrum:
\[
\epsilon(\mathbf{k})=\pm\sqrt{v_1^2(k_x^2+k_y^2)+v_z^2k_z^2}
\] 
Each branch of this spectrum is two fold Kramers degenerate. 

In the similar way it is possible to include the magnetic field in this Hamiltonian.  The field dependent part of the Hamiltonian has two independent real constants $g_1$ and $g_2$ as well as one complex constant $g_3$, which play the role of the $\mathbf{k}$-independent g-factors. Therefore, the total Hamiltonian has the form:
\begin{eqnarray}
H &=& H_k+g_1\tau_z\otimes(H_y\sigma_z+H_x\sigma_x)\nonumber\\
& &+g_2 H_z \tau_0\otimes\sigma_y +g_3\tau_+\otimes(H_x\sigma_z+H_y\sigma_x)\\
& &+g_3^*\tau_-\otimes(H_x\sigma_z+H_y\sigma_x)
\nonumber
\end{eqnarray}
where $\tau_{\pm}=(\tau_x\pm i\tau_y)/2$. This Hamiltonian describes the electronic spectrum of GdB$_4$ near $M$ and $A$ points of the BZ in the lowest linear order in $k_i$ and $H_i$. The spectrum may be calculated analytically and is represented by the expression:
\begin{eqnarray}
& &\epsilon(\mathbf{k},\mathbf{H})= \pm\Bigl[A_{\mathbf{k},\mathbf{H}}\nonumber\\
& &\pm\sqrt{B_{\mathbf{k},\mathbf{H}}+C_{\mathbf{k},\mathbf{H}}+D_{\mathbf{k},\mathbf{H}}+E_{\mathbf{k},\mathbf{H}}}\Bigr]^{1/2}
\end{eqnarray}
where
\[
A_{\mathbf{k},\mathbf{H}}=v_1^2(k_x^2+k_y^2)+v_2^2k_z^2+(g_1^2+|g_3|^2)(H_x^2+H_y^2)+g_2^2H_z^2
\]
\[
B_{\mathbf{k},\mathbf{H}}=4(v_1g_1(k_xH_x+k_yH_y)+v_2g_2k_zH_z)^2
\]
\[
C_{\mathbf{k},\mathbf{H}}=4|g_3|^2H_x^2(g_1^2H_x^2+v_1^2k_y^2++v_2^2k_z^2)
\]
\[
D_{\mathbf{k},\mathbf{H}}=4|g_3|^2H_y^2(g_1^2H_y^2+v_1^2k_x^2++v_2^2k_z^2)
\]
\[
E_{\mathbf{k},\mathbf{H}}=8|g_3|^2H_xH_y(v_1^2k_xk_y-g_1^2H_xH_y)
\]
This spectrum is derived in the lowest order in $k_i$ and $H_i$. Note, that there are few invariants which are proportional to the products $k_ik_j$ and $k_iH_j$. These invariants are irrelevant in the vicinity of $k=0$ point, but may have effect at finite $\mathbf{k}$. Therefore, for the discussion of the splitting of a Dirac point in the finite magnetic field these invariants should be included in the Hamiltonian. 

In the case when the field is directed along z-axis i.e perpendicular to all magnetization vectors the Dirac point is unstable (Fig.2a). As in the case of nonmagnetic materials \cite{Young}, two Weyl points with opposite Chern numbers appear at the points $(0,0,\pm g_2H_z/v_2)$. The only second order invariant which influences the details of the spectrum is $k_zH_z\tau_z\otimes\sigma_0$. It is clear that this term in the Hamiltonian shifts slightly the Weyl points in energy but does not affect the position of the Weyl points. The Kramers degeneracy is lifted everywhere except  the plane $k_z=0$ in linear approximation (Eq. (6)). The second order invariants lift the degeneracy everywhere except exactly $M$ and $A$ points of the BZ (Fig. 2a) in agreement with Ref.\cite{Ramaz1,Ramaz2}.  In these points the Kramers degeneracy is preserved, i.e. the energy of the spin up and spin down electrons is the same. 

When the field is parallel to the magnetization in two sublattices and perpendicular to the magnetization in two other sublattices $\mathbf{H}=(H,\pm H,0)$ in linear approximation (Eq.(6)) a Dirac point splits into two Weyl points with opposite Chern numbers at the points $\sqrt{g_1^2+|g_3|^2}H/v_1(1,\pm 1,0)$. 
In that case the important invariant of the second order is $(k_x k_y)(\beta\tau_+\otimes\sigma_y+h.c.)$, $\beta$ is a complex constant. Due to this invariant, as in the case of nonmagnetic materials \cite{Young}, a small gap proportional to $H^2$ opens in the spectrum. The system becomes insulating. The Kramers degeneracy in linear approximation (Eq.(6)) is preserved on the line determined by the conditions $k_z=0$ and $k_x\pm k_y=0$. Second order invariants lift this degeneracy everywhere except $M$ and $A$ points of the BZ \cite{Ramaz1,Ramaz2}.

The situation is different when the field is directed along x- or y-axis. The field is not orthogonal to any sublattice magnetization. A Dirac point is also unstable. In linear approximation (Eq.(6)) it splits to two Weyl points at the points $(\pm\sqrt{g_1^2-|g_3|^2}H_x/v_1,0,0)$ if $g_1>|g_3|$ and to the nodal circle which is determined by the equation $v_1^2k_y^2+v_2^2k_z^2=(|g_3|^2-g_1^2)H_x^2$ if $g_1<|g_3|$. But higher order invariant $(k_xH_x-k_yH_y)(\gamma\tau_+\otimes\sigma_0+h.c.)$ ($\gamma$ is a complex constant) is not zero at $k_x=0$ or $k_y=0$ and leads to the opening of a small gap in the spectrum (Fig.2b). This gap is quadratic in the field ($\propto H_x^2$). Therefore, for this direction of the field the system is insulating. For that case the Kramers degeneracy is lifted everywhere (Fig.2b), i.e at every point of the BZ the spectrum shows the Zeeman splitting. 
\begin{figure}[tb]
\includegraphics[angle=0,width=1.2\linewidth]{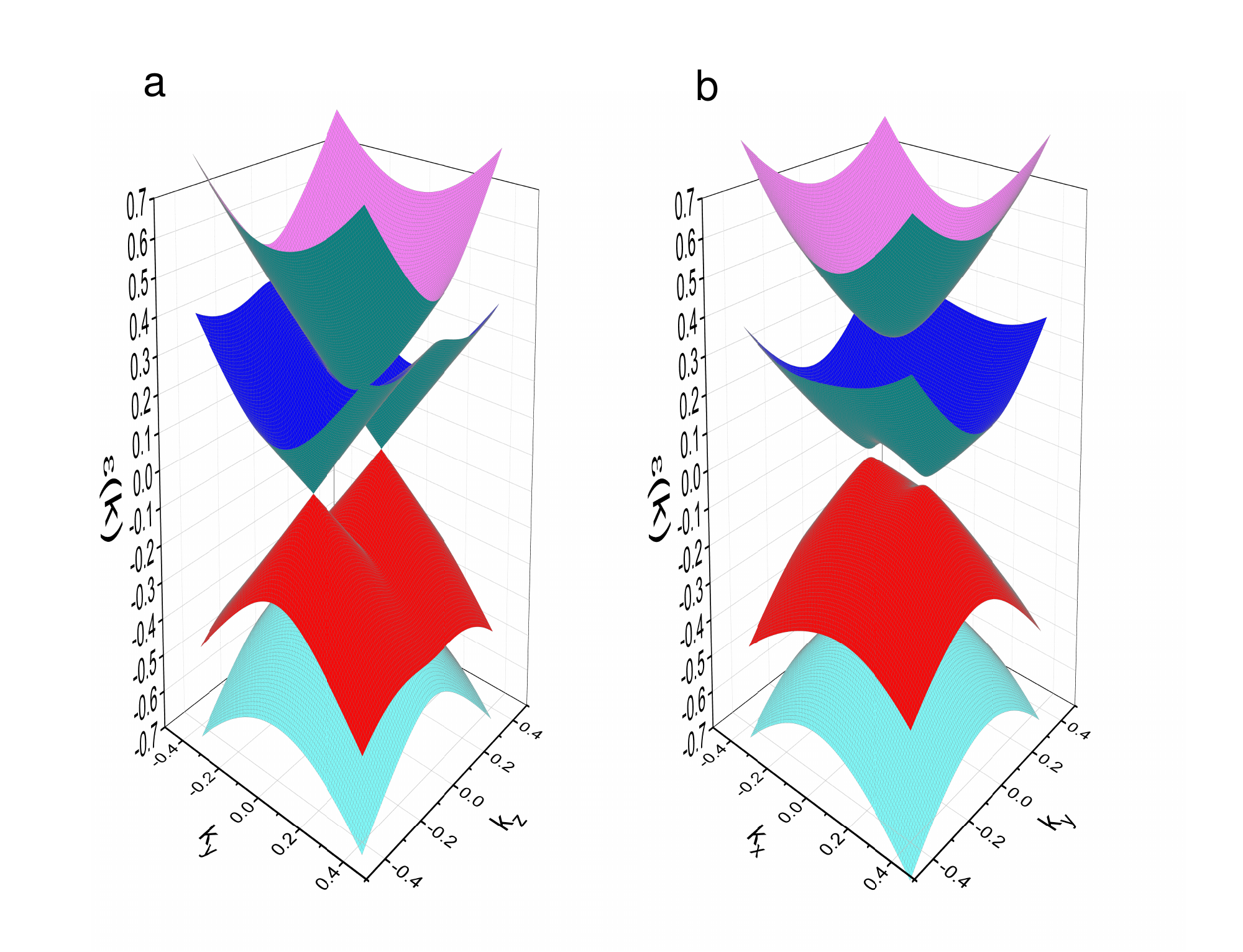}
\caption{a) The splittings of the Dirac point into two Weyl points in the external magnetic field $\mathbf{H}=(0,0,0.2)$ for $v_1=v_2=1$ and $g_1=g_2=1$ and $g_3=1/\sqrt{2}$. All coefficients for higher order invariants are equal to 1. The Kramers degeneracy is preserved at $k_x=k_y=k_z=0$. b) Opening of the gap in the spectrum in the external magnetic field $\mathbf{H}=(0.2,0,0)$. All parameters are the same as in Fig.2a. The gap is quadratic in field. The Kramesrs degeneracy is lifted.}
\end{figure}

\section{Discussion}
Fig. 2 demonstrates how a Dirac spectrum may be controlled by the external magnetic field. Note, that external magnetic field in arbitrary direction eliminates nodal spectrum leading to the small quadratic in the field gap. The gapless spectrum exists only if the external field is orthogonal to the sublattice magnetization. There is another way to induce the true metal-insulating transition in an AFM Dirac semimetal. Since the Dirac point is enforced by the symmetry, the reduction of the four fold rotation symmetry of the Hamiltonian by  any perturbation will lead to the true dielectrization of the spectrum. This reduction of the symmetry may be achieved either by rotation of sublattice magnetization or by applying the symmetry breaking deformation. Indeed, let us consider the deformation of the crystal which is characterized by the strain tensor $\epsilon_{xy}$ and which breaks the four fold rotation axis. The invariant Hamiltonian in that case can be written in the form:
\[
H=H_k+{\cal A}\tau_+\sigma_y\epsilon_{xy}+{\cal A}^*\tau_-\sigma_y\epsilon_{xy},
\] 
here ${\cal A}$ is the complex constant. This leads to the gapped spectrum 
\[
\epsilon(\mathbf{k})=\pm\sqrt{v_1^2(k_x^2+k_y^2)+v_z^2k_z^2+|{\cal A}|^2\epsilon_{xy}^2}.
\] 
This expression describes two branches of the spectra. Each branch is two fold Kramers degenerate. This expression demonstrates that the spectrum of the symmetry enforced Dirac point in the AFM state may be controlled by the perturbation which breaks the four fold rotation axis. 

The main question is whether a Dirac point in $M$ and $A$ high symmetry points of the BZ is in the vicinity of the Fermi energy. There is only one paper with band structure calculations in the paramegnetic phase of GdB$_4$ \cite{Baranovskiy}. From these calculations it is clear that there are branches of the spectrum at $M$ point of the BZ, which are very close to the Fermi energy. On the other hand the spectrum in $A$ point is gapped. 
Therefore, if the formation of the AFM state will not renormalize the spectrum near $M$ point we can expect that the Dirac point will be in close vicinity to the Fermi level.

In conclusion using the symmetry arguments I have shown that the symmetry enforced Dirac points exist in some TRIM of the BZ in GdB$_4$. These Dirac points are very sensitive to external perturbations. The application of the magnetic field leads to disappearance of a Dirac point and appearance of two Weyl points in the spectrum when the field is parallel to z axis and is perpendicular to the magnetization in all sublattices. For other directions of the magnetic field the spectrum is gapped. Dirac points also disappear when the AFM order is destroyed. And finally, the application of the perturbation which breaks the four fold axis in the AFM state leads to the gapped spectrum. 

\section{Acknowledgments}
I thank Y. Gerasimenko and V. Nasretdinova for very helpful discussions. I also acknowledge the financial support from Slovenian Research Agency Program P1-0040.

\end{document}